\documentclass{aa}

\usepackage{graphicx}
\usepackage{rotating}

\def\i{\,{\sc i}}
\def\ii{\,{\sc ii}}

\def\SM{{\sc SynthM}}
\def\atlas9{{\sc ATLAS9}}

\begin{document}

\title{
On the influence of Stark broadening on Cr\i\ lines in
stellar atmospheres}

\author{M. S. Dimitrijevi\'c\inst{1,2} \and T. Ryabchikova\inst{3,4} \and L. \v C. Popovi\'c\inst{1,2,5} \and D. Shulyak\inst{6} \and S. Khan\inst{6}}

\institute{Astronomical Observatory, Volgina 7, 11160 Belgrade 74,
Serbia \and
 Isaac Newton Institute of Chile, Yugoslavia Branch
\and
Institute of Astronomy, Russian Academy of Science, Pyatnitskaya 48,
119017 Moscow, Russia
\and
Institute for Astronomy, University of Vienna, T\"urkenschanzstrasse 17,
A-1180 Vienna, Austria
\and
Astrophysikalisches Institut Potsdam, An der Sternwarte 16, 14482
Potsdam, Germany
\and
Tavrian National University, Yaltinskaya 4, 330000 Simferopol, Crimea,
Ukraine}

\date{Received  / Accepted }

\offprints{M. S. Dimitrijevi\'c}
\authorrunning{M. S. Dimitrijevi\'c et al.}
\titlerunning{Stark broadening of Cr\i\ lines}

\abstract{Using the semiclassical perturbation method , electron-,
proton-, and ionized helium-impact line widths and shifts for the
nine Cr\i\ spectral lines from the $4p^7P^0-4d^7D$ multiplet, have
been calculated for a perturbers density of 10$^{14}$ cm$^{-3}$
and temperatures T $=$ 2,500 $-$ 50,000 K. The obtained results
have been used to investigate the influence of Stark broadening
effect in the Cr-rich Ap star $\beta $ CrB atmosphere on line
shapes of these lines. It has been found that the contribution of
the proton and He\ii\ collisions to the line width and shift is
significant, and it is comparable and sometimes (depending of the
electron temperature) even larger than electron-impact
contribution. Moreover, not only the Stark line width, but,
depending on the electron-, proton-, and He\ii\ density in stellar
atmosphere also the Stark shift may contribute to the blue as well
as to the red asymmetry of the same line. The obtained results
have been used to investigate the influence of Stark broadening
effect on line shapes of Cr\i\, lines in the atmosphere of Cr-rich
Ap star $\beta $ CrB. }

\maketitle

\section{Introduction}

The Stark broadening is the most significant pressure broadening
mechanism  for A and B stars and this effect should be  taken into
account in investigation, analysis and modelling of their
atmospheres. In our previous works (Popovi\'c et al. 1999,
Popovi\'c et al. 2001, Dimitrijevi\'c et al. 2003) we have shown
that the Stark broadening may change the spectral line equivalent
widths
 by 10-45\%, hence neglecting this mechanism,
a significant errors in abundance determinations may be
introduced. Stark data become extremely important after
discovering abundance gradients in the atmospheres of magnetic
peculiar (Ap) stars (Babel 1992, Ryabchikova et al. 2002, Wade et
al. 2003). High resolution spectra allow us to perform
stratification analysis using line profiles, and strong lines with
developed wings provide us with the most accurate information
about distribution of the element through the stellar atmosphere
(see Dimitrijevi\'c et al. 2003 for Si).

Chromium is one of the most anomalous elements in Ap stars. It was
shown to be concentrated in the deeper atmospheric layers in Ap
stars $\beta$~CrB (Wade et al. 2003) and in $\gamma$~Equ
(Ryabchikova et al. 2002), where electron density is high enough
to favour the Stark effect. The most of Cr\i\, and Cr\ii\, lines
in the optical spectral region have rather small Stark damping
constants and hence, no measurable Stark wings appeared. However,
Cr\i\, lines from $4p-4d$ transitions are known to have rather
large Stark damping constants according to calculations made by
Kurucz (1993).

We present new calculations of Cr \i\ lines Stark widths and
shifts based on the semiclassical perturbation approach
(Sahal-Br\'echot 1969a,b, 1974, Dimitrijevi\'c \& Sahal-Br\'echot
1984a).  Method of these calculations is described in
Section~\ref{semiclass}, magnetic synthetic spectrum calculations
are described in Section~\ref{synthM}. Results of Stark broadening
calculations are given in Section~\ref{broad}. An application of
new data to spectrum synthesis in magnetic Ap star $\beta$~CrB is
considered in Section~\ref{CrB}.


\section{The Stark broadening parameter calculation} \label{semiclass}

Calculations have been performed within the semiclassical perturbation
formalism,  developed
and discussed in detail in Sahal-Br\'echot (1969ab). This formalism,
as well as the
corresponding computer code, have been
optimized and updated several times (see e.g. Sahal-Br\'echot,
1974; Dimitrijevi\'c and Sahal-
Br\'echot, 1984a, Dimitrijevi\'c, 1996).

Within this formalism, the
full width of a neutral emitter isolated spectral line broadened by
electron impacts can be expressed in
terms of cross sections for elastic and inelastic processes as

\begin{equation}
W_{if} = {2\lambda_{if}^2\over 2\pi c}n_e\int vf(v)dv (\sum_{i'\ne
i}\sigma_{ii'}(v) + \sum_{f'\ne f}\sigma_{ff'}(v) + \sigma_{el})
\end{equation}
\noindent{and the corresponding line shift as}

\begin{equation}
d_{if} = {\lambda_{if}^2\over 2\pi c}n_e\int vf(v)dv\int_{R_3}^{R_D}
2\pi \rho
d\rho \sin 2\phi_p
\end{equation}

\noindent{Here, $\lambda_{if}$ is the wavelength of the line
originating from the transition with the initial atomic energy level
$i$ and the final level $f$, $c$ is the velocity of light, $n_e$ is the
electron density, $f(v)$ is the Maxwellian velocity distribution
function for electrons, $m$ is the electron mass, $k$ is the Boltzmann constant,
$T$ is
the temperature, and
$\rho $ denotes the impact parameter of the
incoming electron. The inelastic cross section $\sigma_{jj'}(v)$ is
determined according to Chapter 3
in Sahal-Br\'echot (1969b), and elastic cross section $\sigma_{el}$ according
to Sahal-Br\'echot (1969a). The cut-offs, included in
order to maintain for the unitarity of the {\bf $S$}-matrix, are
described in Section 1 of Chapter 3 in Sahal-Br\'echot (1969a).}

The formulae for the ion-impact broadening parameters are
analogous to the formulae for electron-impact broadening. For the
colliding ions, the validity of the impact approximation has to be
checked in the far wings.

\section{Line profile calculations} \label{synthM}

Model atmosphere calculations as well as calculations of the absorption
coefficients
were made with the local thermodynamical equilibrium (LTE) approximation.
Model calculations were performed with the ATLAS9 code (Kurucz
1993), modified by Piskunov \& Kupka (2001) to include individual chemical
composition of the star in line opacity calculations (see Kupka et al. 2004).

 The next step is the calculation of the outward flux at corresponding
wavelengths points using the given model.  For this purpose we
used \SM\, code written by Khan (2004).
 This code allows to calculate synthetic spectra of early and intermediate type of stars taking into account
 magnetic field effects and stratification of chemical elements.

The computational scheme  is as follows. For each line we find the
central opacity as

\begin{equation} \label{eq:a1}
\alpha_{line}=\frac{\pi e^2}{mc} gf_{if}
e^{-\frac{\chi}{kT}}\frac{n}{\rho}(1-e^{-\frac{h\nu}{kT}}),
\end{equation}

\noindent{where $\alpha_\nu$ is the mass absorption coefficient at
frequency $\nu$, $e$ is the electron charge, $g$ is the
statistical weight, $f_{if}$ is the oscillator strength for a
given transition, $\chi$ is the excitation energy,
 $n$ is the number density of a corresponding element in a given
ionization stage multiplied by partition function, $\rho$ is the
density and $h$ is the Planck constant. The last factor describes
stimulated emission.}

Next, we compute the total damping parameter

\begin{equation}
\gamma = \gamma_{rad} + \gamma_{Stark} + \gamma_{neutral}.
\end{equation}

\noindent{Here $\gamma_{rad}$,  $\gamma_{Stark}$ and
$\gamma_{neutral}$ are the radiative broadening, Stark broadening
and damping parameters due to neutral atom collisions
respectively. The values of $\gamma_{rad}$, $\gamma_{neutral}$,
excitation energy $\chi$ and oscillator strength $gf$ were taken
from the Vienna Atomic Line Database (VALD) (Kupka et al. 1999).
In the case of neutral atom broadening we assumed that perturbing
particles are  atoms of neutral hydrogen and helium only (Van der
Waals broadening). This assumption is applicable to almost all
types of stars due to high hydrogen and helium cosmic abundances.
We shall discuss the competition between the broadenings caused by
the Stark effect and neutral hydrogen collisions in the
atmospheres of our template star $\beta$~CrB in Section 4.2.

In order to include Stark broadening effects we added the
approximate formulas (see Eqs. (20) and (21) in Sec. 4.1) in the
code. The Stark width and shift are

\begin{equation}
\gamma_{Stark} =  \gamma_{Stark}^{(e)}n_e + \gamma_{Stark}^{(p)}n_p +
\gamma_{Stark}^{(HeII)}n_{HeII},
\end{equation}

\begin{equation}
d_{Stark} =  d_{Stark}^{(e)}n_e + d_{Stark}^{(p)}n_p +
d_{Stark}^{(HeII)}n_{HeII},
\end{equation}
where $n_e$, $n_p$ and $n_{HeII}$  are the corresponding densities of
electrons,
protons and
He\ii\, ions respectively. The  resulting opacity profile
is given by the Voigt function (Doppler + pressure broadening).

The next step is to solve transfer equation with new Stark damping
parameters. In presence of magnetic field atomic level $k$
determined by quantum numbers $J_k$, $L_k$, $S_k$ splits into
$2J_k+1$ states with $M_k=-J_k,\dots,+J_k$. The absolute value of
splitting is defined by field modulus $|\emph{\textbf{B}}|$ and
Land\'{e} factor $g_k$ which in the case of LS coupling is
calculated as
\begin{equation} \label{lande_factor}
g_k=\frac{3}{2}+\frac{S_k(S_k+1)-L_k(L_k+1)}{2J_k(J_k+1)}.
\end{equation}
The possible transitions between splitted levels ($u$ - upper and
$l$ - lower level) are allowed by selection rules
\begin{equation} \label{selection_rule}
\Delta{M}=M_u-M_l=\left\{
\begin{array}{r}
+1\equiv b,\\0\equiv p,\\-1\equiv r.\\
\end{array}
\right.
\end{equation}
For a normal Zeeman triplet the subscript $p$ corresponds to the
unshifted $\pi$ component, whereas $b$ and $r$ correspond to the
blue- and red-shifted $\sigma$ components respectively. In the
general case of an anomalous splitting indices $p$, $b$, $r$ refer
to the series of the $\pi$ and $\sigma$ components.

The wavelength shift of the component relative to the laboratory
line centre $\lambda_0$ is defined by
\begin{equation} \label{delta_nu_zeeman}
\Delta\lambda=\frac{e\lambda_0^2|\vec{B}|}{4\pi
mc^2}\,(g_lM_l-g_uM_u)\,.
\end{equation}
The relative strengths of components $S_{p,r,b}$ in accordance
with Sobelman (1977) are proportional to
\begin{equation}
\left(
\begin{array}{ccc}
J_u & 1 & J_l \\
-M_u & M_u-M_l & M_l
\end{array}
\right)^2,
\end{equation}
where the last structure is a $3j$-symbol. The normalization of
components is done so that
\begin{equation} \label{gf_normalization}
\sum S_p=\sum S_b=\sum S_r=1.
\end{equation}

In general case the polarized radiation can be described by means
of Stokes $IQUV$ parameters. The transfer equation is
\begin{equation} \label{transfer_equation}
\frac{d\vec{I}}{dz}=-\vec{K}\vec{I}+\vec{J}.
\end{equation}
Here $\vec{I}=(I,Q,U,V)^{T}$ is the Stokes vector, $\vec{K}$ is
the absorption matrix and $\vec{J}$ is the emission vector,
\begin{equation} \label{alpha_tot}
\vec{K}=\alpha_{c}\vec{1}+\sum_{lines}\alpha_{line}\vec{\Phi}_{line},
\end{equation}
\begin{equation} \label{emission_vector}
\vec{J}_\nu=\alpha_{c}B_{\nu}(T)\vec{e_{0}}+B_{\nu}(T)\sum_{lines}\alpha_{line}\vec{\Phi}_{line}\vec{e_0},
\end{equation}
where $\vec{1}$ is the ${4\times4}$ unit matrix,
$\vec{e_0}=(1,0,0,0)^{T}$, $\alpha_{c}$ is the continuum
absorption coefficient, $\alpha_{line}$ is the line central
opacity for zero damping and zero magnetic field given by equation
(3). Here we also assume LTE so the source function is equal to
Planck function $B_{\nu}(T)$.

The line absorption matrix $\vec{\Phi}_{line}$ is created using
absorption profiles $\phi_j$ and anomalous dispersion profiles
$\psi_j$ of $\pi$ and $\sigma_\pm$ components in accordance with
Rees et al. (1989). These profiles are calculated as (for
$j=p,b,r$)
\begin{equation} \label{phi}
\phi_{j}=\sum_{i_j=1}^{N_j}S_{i_j}H\left(a,\upsilon-\frac{\Delta\lambda_{i_j}}{\Delta\lambda_D}\right),
\end{equation}
\begin{equation} \label{psi}
\psi_{j}=2\sum_{i_j=1}^{N_j}S_{i_j}F\left(a,\upsilon-\frac{\Delta\lambda_{i_j}}{\Delta\lambda_D}\right),
\end{equation}
where $H(a,u)$ and $F(a,u)$ are the Voigt and Faraday-Voigt
functions,
\begin{equation} \label{a_damping}
a=\frac{\gamma\,\lambda_0^2}{4\pi c\,\Delta\lambda_{D}},
\end{equation}
\begin{equation} \label{u_damping}
\upsilon=\frac{\lambda-\lambda_0+d_{Stark}}{\Delta\lambda_D}.
\end{equation}
The Stark shift $d_{Stark}$ and the damping parameter $\gamma$
have been found from (6) and (4), respectively. The Doppler width
$\Delta\lambda_D$ is
\begin{equation} \label{doppler_nu}
\Delta\lambda_D = \frac{\lambda}{c}\sqrt{\frac{2kT}{m_A}+\xi^2_t},
\end{equation}
where $m_A$ is the mass of the absorber atom and $\xi_t$ is the
microturbulent velocity.

The transfer equation (\ref{transfer_equation}) is solved by DELO
method (Rees et al. 1989) with quadratic approximation for the
source function which was suggested by Socas-Navarrro et al.
(2000) and written in Fortran by Piskunov \& Kochukhov (2000).
Also, we used fast Humlicek (1982) algorithm for approximations of
Voigt and Faraday-Voigt functions.

The convolution of the synthetic spectra due to rotation of the
star was produced by IDL procedure of Valenti \& Anderson
implemented in the {\sc rotate} IDL code by N.~Piskunov.

\section{Results}

\subsection{Stark broadening data} \label{broad}

The atomic energy levels needed for Stark broadening calculations
were taken from
Wiese and Musgrove (1989).
Oscillator strengths have been
calculated by using the method of Bates and Damgaard (1949) and
tables of
Oertel and Shomo (1968). For
higher levels, the
method described in van Regemorter, Binh Dy and Prud'homme (1979) has
been applied.

The spectrum of neutral chromium is complex and not known well
enough for a good calculation of the considered lines. The
principal problem  is the absence of the reliable experimental
data for $f$ levels. However, by inspecting theoretically
predicted Cr I levels in Moore (1971), one can see that the $4f$
level is missing. In accordance with the decrease of distance
between existing $s,\- p,$\- and $d$ levels, we estimated that the
eventual 5$f$ level should be around 52000 cm$^{-1}$ and not
closer than 50000 cm$^{-1}$. We checked results without 5$f$ level
and with a fictive 5$f$ level at 52000 and 50000 cm$^{- 1}$. In
all cases, line widths differed by less than 1\%, while the shift
varies within the limits of several percents. Consequently we
performed calculations without the contribution of $f$ energy
levels. Since the average estimated error of the semiclassical
method is $\pm $30\%, due to additional approximations and
uncertainties, we estimate the error bars of our results to be
$\pm $50\%.

Our results for electron-, proton-, and
ionized helium-impact line widths and shifts for the nine considered Cr\i\
spectral
lines for a perturbed density of 10$^{14}$ cm$^{-3}$
and temperatures T $=$ 2,500 $-$ 50,000 K, are shown in Table 1.
For
perturber densities
lower than those tabulated here, Stark broadening parameters vary
linearly with perturber density. The nonlinear behavior of Stark broadening
parameters
 at higher densities is the consequence of the influence of Debye
shielding and has been analyzed in detail in Dimitrijevi\'c and
Sahal-Br\'echot (1984b).

 \begin{table*}
\begin{center}
      \caption[]{ Stark broadening parameters for Cr\i\,
$4p$ - $4d$ spectral lines. This table shows electron-, proton-,
and ionized helium-impact broadening parameters for Cr\i\, for a
perturber density of  10$^{14}$ cm$^{-3}$ and temperatures from
2,500 up to 50,000 K. The quantity  C (given in \AA\- cm$^{-3}$),
when divided by the corresponding full width at half maximum,
gives an estimate for the maximum perturber density for which
tabulated data may be used. For higher densities, the isolated
line approximation used in the calculations breaks down.The
asterisk identifies cases for which the collision volume
multiplied by the perturber density (the condition for validity of
the impact approximation) lies between 0.1 and 0.5. When this
value is larger than 0.5, the corresponding results have been
omitted. FWHM(\AA) denotes full line width at half maximum in \AA
, while SHIFT(\AA) denotes line shift in \AA.} \label{Stark_1}
\begin{tabular}{|r|r|r|r|r|r|r|r|}
\hline
\multicolumn{2}{|c|}{PERTURBERS ARE:}& \multicolumn{2}{c|}{ELECTRONS}& \multicolumn{2}{c|}{
PROTONS}&\multicolumn{2}{c|}{HELIUM IONS} \\
\hline
TRANSITION    &          T(K)  &  FWHM(A)  &   SHIFT(A) &   FWHM(A)  &   SHIFT(A)  &  FWHM(A)   &
SHIFT(A) \\
\hline
 Cr I  $^7$P$^o_2$ - $^7$D$_1$&  2500. & 0.890E-02& -0.205E-02&  0.461E-02& -0.379E-02 & 0.344E-02& -0.299E-02\\
     &  5000. & 0.772E-02& -0.146E-02&  0.551E-02& -0.437E-02 & 0.388E-02& -0.346E-02\\
   5276.07 A  & 10000. & 0.655E-02& -0.104E-02&  0.678E-02& -0.505E-02 & 0.439E-02& -0.396E-02\\
 C = 0.43E+15 & 20000. & 0.548E-02& -0.767E-03&  0.832E-02& -0.592E-02 & 0.503E-02& -0.451E-02\\
             & 30000. & 0.493E-02& -0.660E-03&  0.915E-02& -0.651E-02 & 0.547E-02& -0.488E-02\\
             & 50000. & 0.431E-02& -0.559E-03&  0.993E-02& -0.719E-02 & 0.602E-02& -0.544E-02\\
\hline
 Cr I  $^7$P$^o_2$ - $^7$D$_2$&  2500. & 0.193E-01& -0.216E-02&&&&\\
     &  5000. & 0.164E-01& -0.155E-02&&&&\\
   5275.75 A  & 10000. & 0.136E-01& -0.108E-02&&&&\\
 C = 0.11E+14 & 20000. & 0.111E-01& -0.790E-03& *0.717E-01& -0.298E-01&&\\
             & 30000. & 0.983E-02& -0.675E-03& *0.702E-01& -0.273E-01&&\\
             & 50000. & 0.841E-02& -0.569E-03& *0.664E-01& -0.239E-01& *0.135E-01& -0.346E-01\\
\hline
 Cr I  $^7$P$^o_2$ - $^7$D$_3$&  2500. & 0.240E-01& -0.939E-03& *0.216E-01& -0.138E-01& *0.131E-01& -0.102E-01\\
     &  5000. & 0.210E-01& -0.677E-03& *0.253E-01& -0.170E-01& *0.151E-01& -0.124E-01\\
   5275.28 A  & 10000. & 0.178E-01& -0.473E-03& *0.274E-01& -0.193E-01& *0.169E-01& -0.150E-01\\
 C = 0.60E+14 & 20000. & 0.147E-01& -0.401E-03&  0.280E-01& -0.193E-01& *0.170E-01& -0.179E-01\\
             & 30000. & 0.131E-01& -0.393E-03&  0.283E-01& -0.180E-01& *0.157E-01& -0.193E-01\\
             & 50000. & 0.112E-01& -0.393E-03&  0.289E-01& -0.155E-01& *0.127E-01& -0.200E-01\\
\hline
 Cr I  $^7$P$^o_3$ - $^7$D$_2$&  2500. & 0.195E-01& -0.219E-02&&&&\\
     &  5000. & 0.165E-01& -0.156E-02&&&&\\
   5298.49 A  & 10000. & 0.137E-01& -0.109E-02&&&&\\
 C = 0.12E+14 & 20000. & 0.112E-01& -0.801E-03& *0.723E-01& -0.301E-01&&\\
             & 30000. & 0.992E-02& -0.685E-03& *0.708E-01& -0.276E-01&&\\
             & 50000. & 0.849E-02& -0.578E-03& *0.670E-01& -0.242E-01& *0.136E-01& -0.349E-01\\
\hline
 Cr I  $^7$P$^o_3$ - $^7$D$_3$&  2500. & 0.243E-01& -0.950E-03& *0.218E-01& -0.140E-01& *0.133E-01& -0.103E-01\\
    &  5000. & 0.212E-01& -0.687E-03& *0.256E-01& -0.171E-01& *0.153E-01& -0.125E-01\\
   5298.02 A  & 10000. & 0.179E-01& -0.481E-03& *0.276E-01& -0.195E-01& *0.171E-01& -0.151E-01\\
 C = 0.60E+14 & 20000. & 0.148E-01& -0.408E-03&  0.283E-01& -0.195E-01& *0.171E-01& -0.180E-01\\
             & 30000. & 0.132E-01& -0.400E-03&  0.286E-01& -0.182E-01& *0.158E-01& -0.195E-01\\
             & 50000. & 0.113E-01& -0.400E-03&  0.291E-01& -0.157E-01& *0.128E-01& -0.201E-01\\
\hline
\hline
\end{tabular}
\end{center}
\end{table*}
\newpage
\addtocounter{table}{-1}
\begin{table*}[t!]
\begin{center}
\caption{(continued)}
\begin{tabular}{|r|r|r|r|r|r|r|r|}
\hline \multicolumn{2}{|c|}{PERTURBERS ARE:}&
\multicolumn{2}{c|}{ELECTRONS}& \multicolumn{2}{c|}{
PROTONS}&\multicolumn{2}{c|}{HELIUM IONS} \\
\hline TRANSITION    &          T(K)  &  FWHM(A)  &   SHIFT(A) &
FWHM(A)  &   SHIFT(A)  &  FWHM(A)   &
SHIFT(A) \\
\hline

 Cr I  $^7$P$^o_3$ - $^7$D$_4$&  2500. & 0.182E-01 & 0.452E-03 & 0.109E-01& -0.737E-02& *0.683E-02& -0.564E-02\\
     &  5000. & 0.156E-01 & 0.277E-03 & 0.130E-01& -0.902E-02& *0.788E-02& -0.664E-02\\
   5297.38 A  & 10000. & 0.131E-01 & 0.129E-03 & 0.144E-01& -0.102E-01&  0.909E-02& -0.787E-02\\
 C = 0.12E+15 & 20000. & 0.108E-01 &-0.434E-04 & 0.168E-01& -0.945E-02&  0.956E-02& -0.934E-02\\
             & 30000. & 0.956E-02 &-0.141E-03 & 0.172E-01& -0.830E-02&  0.945E-02& -0.100E-01\\
             & 50000. & 0.822E-02 &-0.240E-03 & 0.194E-01& -0.603E-02&  0.793E-02& -0.100E-01\\
\hline
 Cr I  $^7$P$^o_4$ - $^7$D$_3$&  2500. & 0.245E-01& -0.956E-03& *0.220E-01& -0.141E-01& *0.134E-01& -0.104E-01\\
     &  5000. & 0.214E-01& -0.701E-03& *0.259E-01& -0.173E-01& *0.155E-01& -0.127E-01\\
   5329.78 A  & 10000. & 0.181E-01& -0.492E-03& *0.280E-01& -0.197E-01& *0.173E-01& -0.153E-01\\
 C = 0.61E+14 & 20000. & 0.150E-01& -0.419E-03&  0.286E-01& -0.197E-01& *0.173E-01& -0.182E-01\\
             & 30000. & 0.134E-01& -0.411E-03&  0.289E-01& -0.184E-01& *0.160E-01& -0.197E-01\\
             & 50000. & 0.115E-01& -0.410E-03&  0.295E-01& -0.159E-01& *0.130E-01& -0.204E-01\\
\hline
 Cr I  $^7$P$^o_4$ - $^7$D$_4$&  2500. & 0.184E-01 & 0.460E-03&  0.110E-01& -0.746E-02& *0.692E-02& -0.570E-02\\
     &  5000. & 0.158E-01 & 0.276E-03&  0.132E-01& -0.911E-02& *0.798E-02& -0.672E-02\\
   5329.14 A  & 10000. & 0.132E-01 & 0.127E-03&  0.146E-01& -0.103E-01&  0.920E-02& -0.796E-02\\
 C = 0.12E+15 & 20000. & 0.109E-01 &-0.484E-04&  0.170E-01& -0.956E-02&  0.967E-02& -0.946E-02\\
             & 30000. & 0.968E-02 &-0.147E-03&  0.174E-01& -0.840E-02&  0.956E-02& -0.101E-01\\
             & 50000. & 0.832E-02 &-0.248E-03&  0.197E-01& -0.610E-02&  0.802E-02& -0.101E-01\\
\hline
 Cr I  $^7$P$^o_4$ - $^7$D$_5$&  2500. & 0.920E-02 & 0.162E-02 & 0.499E-02 & 0.407E-02&  0.370E-02 & 0.321E-02\\
     &  5000. & 0.800E-02 & 0.979E-03 & 0.599E-02 & 0.470E-02&  0.417E-02 & 0.372E-02\\
   5328.32 A  & 10000. & 0.678E-02 & 0.474E-03 & 0.739E-02 & 0.544E-02&  0.473E-02 & 0.426E-02\\
 C = 0.40E+15 & 20000. & 0.567E-02 & 0.103E-03 & 0.902E-02 & 0.641E-02&  0.543E-02 & 0.485E-02\\
             & 30000. & 0.509E-02 &-0.604E-04 & 0.987E-02 & 0.709E-02&  0.590E-02 & 0.526E-02\\
             & 50000. & 0.445E-02 &-0.208E-03 & 0.106E-01 & 0.784E-02&  0.647E-02 & 0.589E-02\\

\hline
\end{tabular}
\end{center}
\end{table*}

\begin{figure}[h!]
\includegraphics[width=8.5cm]{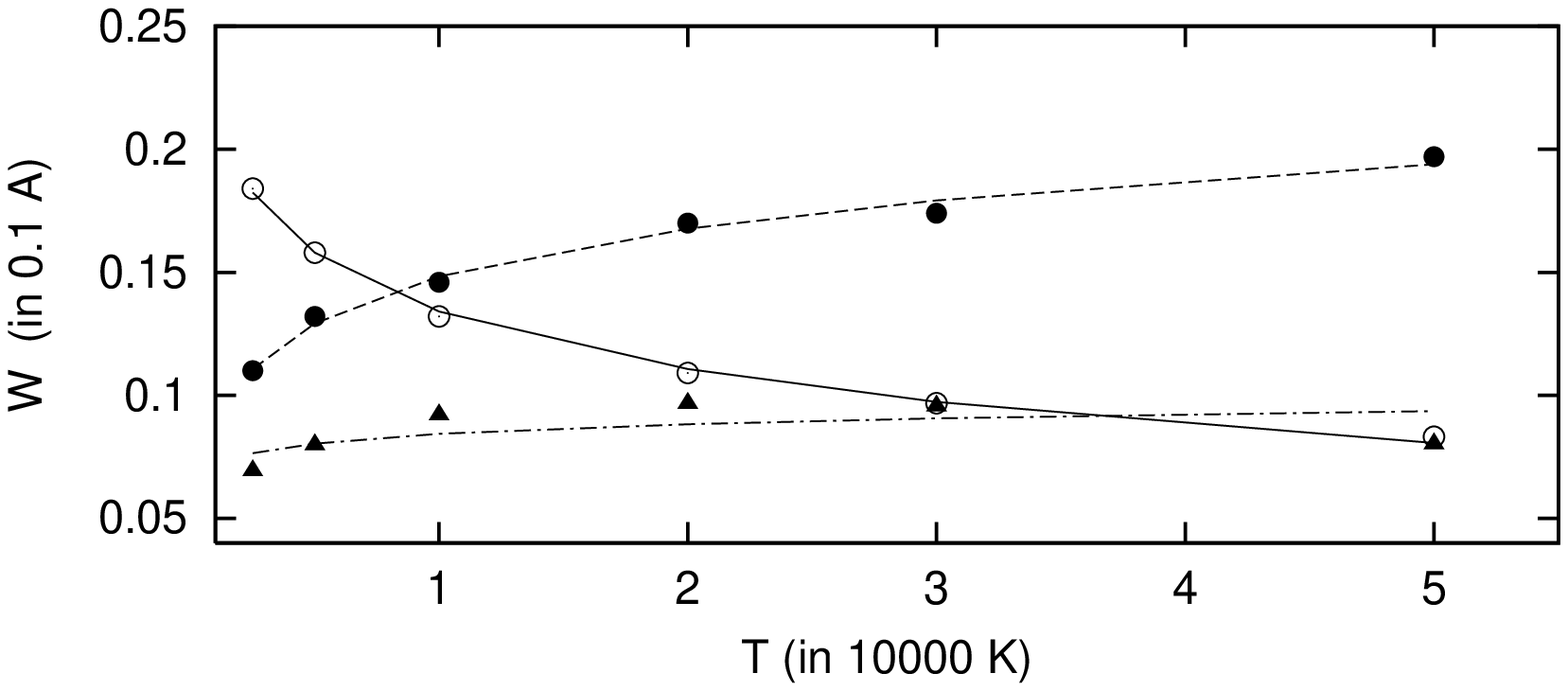}
\includegraphics[width=8.5cm]{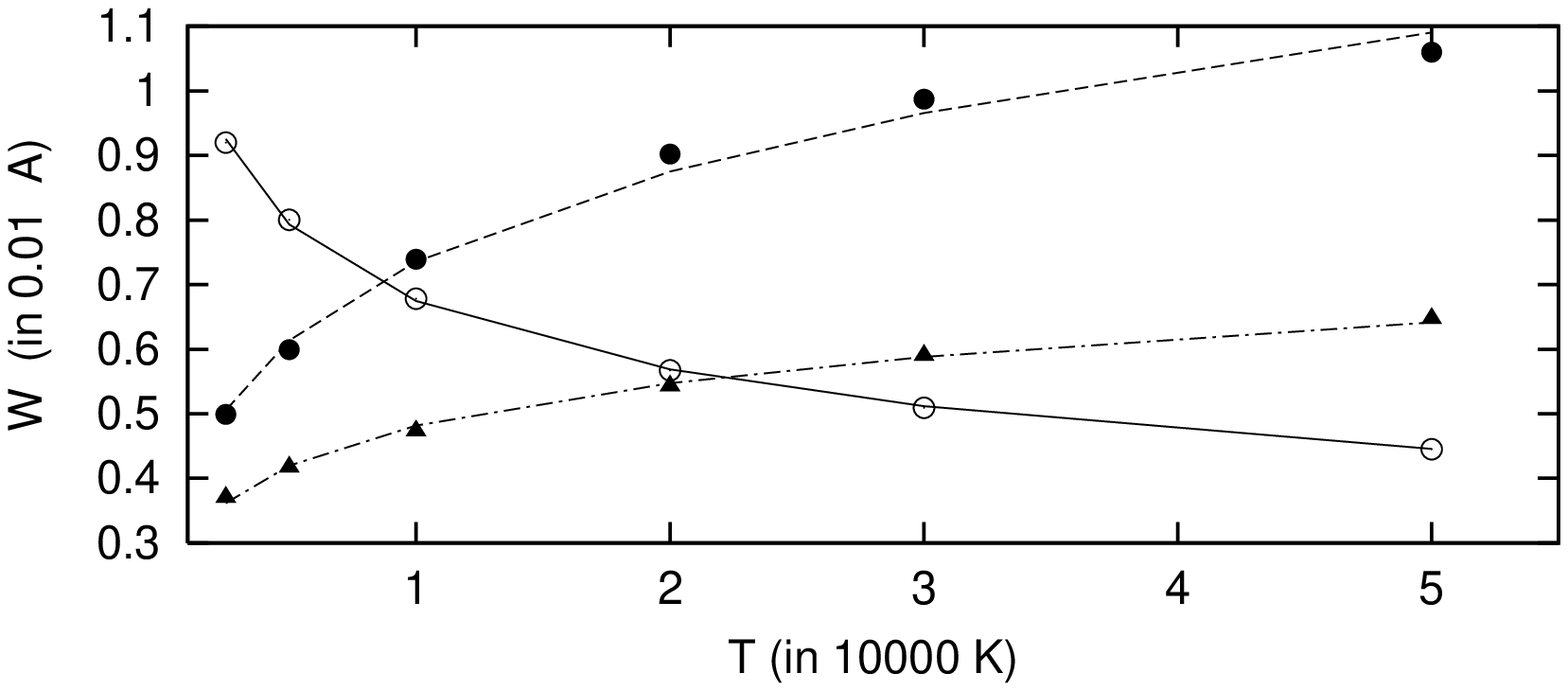}
\caption{The analytic fit  of Cr I $\lambda=5329.14$\ \AA \-
(upper Fig.) and $\lambda=5328.32$ \AA \- (down) Stark widths due
to impact with: electrons (open circles), protons (full circles)
and He\ii\ (full triangles).}
 \label{fig1}
\end{figure}

\begin{figure}[h!]
\includegraphics[width=8.5cm]{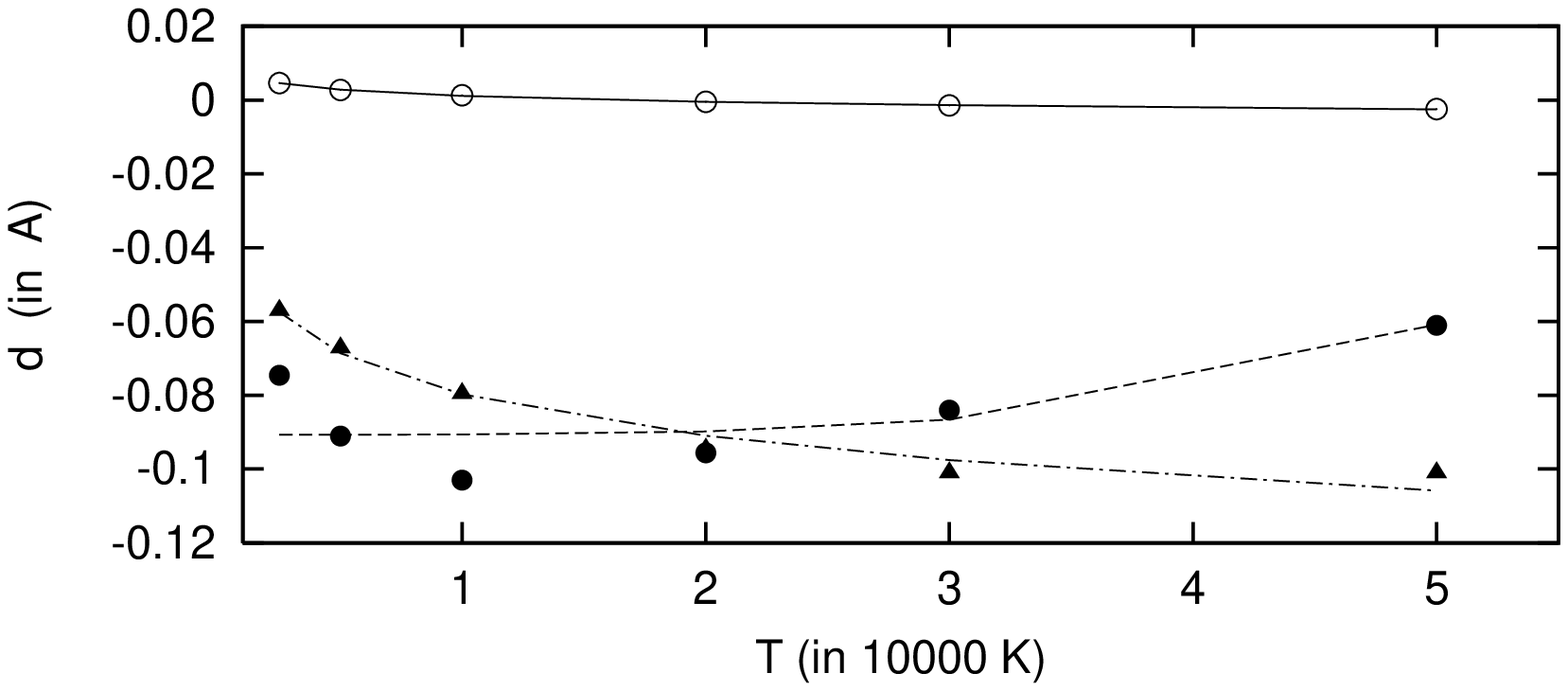}
\includegraphics[width=8.5cm]{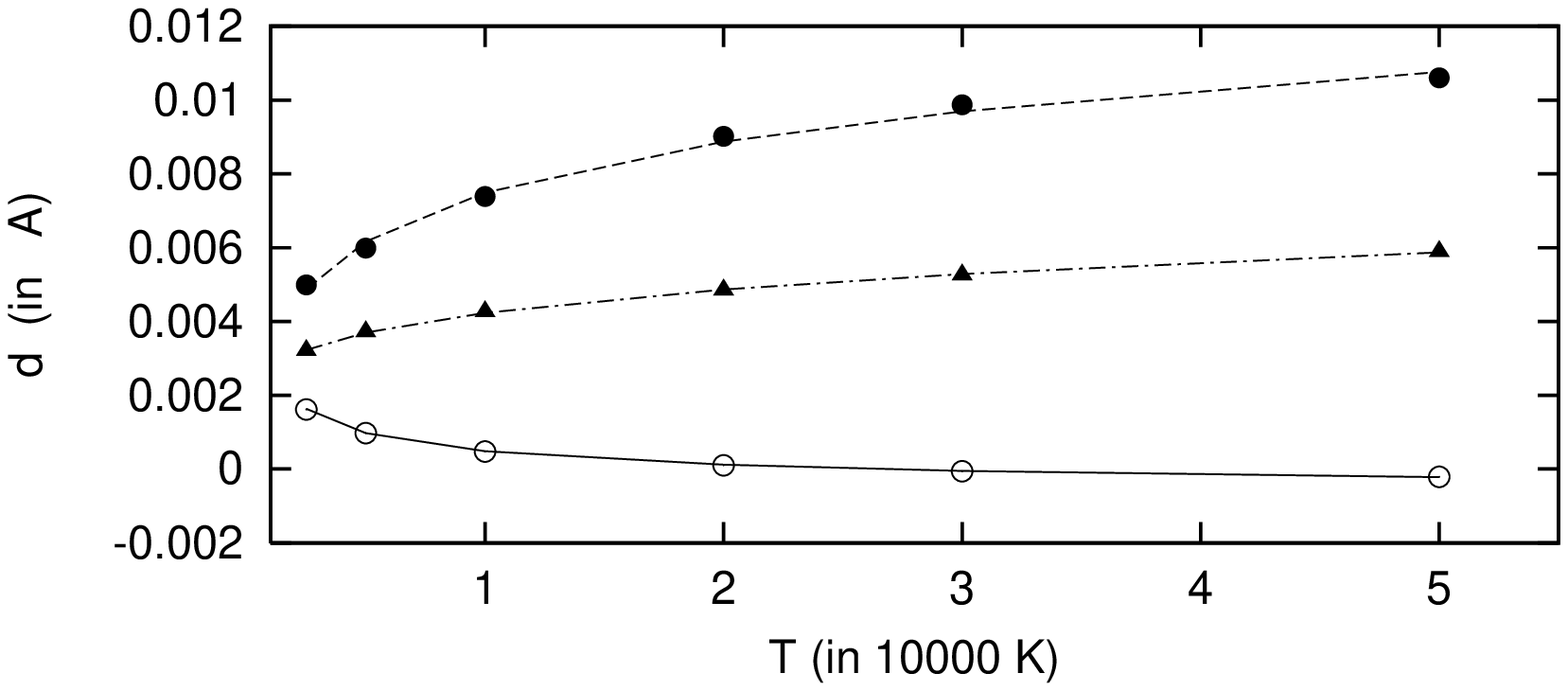}
\caption{The same as in Fig. 1, but for Stark shift.}
 \label{fig1}
\end{figure}


After testing the density dependence
of Stark parameters, we have found that the width and shift are linear functions
of  density for perturber densities smaller than 10$^{16}\rm
cm^{-3}$ and can be scaled by the simple formula:

\begin{equation}
(W,d)_N=(W,d)_0({N\over 10^{14}}),
\end{equation}
where $(W,d)_N$ are the width and shift at a perturber
density $N$
($\rm cm^{-3}$), and
$(W,d)_0$  are width and shift  given in Table 1, respectively.

In order to simplify the use of Stark broadening data in the codes for stellar
spectral synthesis, we have found an analytical expression for Stark
widths and shifts

\begin{equation}
{W\over n_e}{\rm [\AA]}=c_1\cdot(A+T^B),
\end{equation}

\begin{equation}
{d\over n_e}{\rm [\AA]}=c_2\cdot(A+C\cdot T^B).
\end{equation}
The constants $c_1,\ c_2,\ A$, $B$ and $C$ are given in Table 2.
We take $T$  as $T/(10000 K)$. The analytical fit is present in
Figs. 1 and 2. As one can see from the Figs. 1 and 2, the
analytical assumption good fit the calculated value in both, Stark
widths and shifts. Here we should note that in some lines (as e.g.
in Cr I $\lambda=$ 5329.14\ \AA) the analytical fit in the case of
proton and He\ii\ impact broadening parameters not exactly fit the
calculated values (see e.g. Fig. 2, fit for proton-impact shift),
but the differences are smaller than the expected error of our
calculations ($\approx\pm$50\%).

\begin{table*}
\begin{center}
     \caption[]{The parameters A, B and C of the approximate formulae
 for Stark widths and shifts.}
\begin{tabular}{|c|c|c|c|c|c|c|c|c|c|}
\hline
Line &5276.07\ \AA\ & 5275.75\ \AA\ &5275.28\ \AA\ & 5298.49\ \AA\ & 5298.02\
\AA\ & 5297.38\ \AA\ & 5329.78\ \AA\ & 5329.14\ \AA\ & 5328.32\ \AA\ \\
\hline
\hline
\multicolumn{10}{|c|}{WIDTH} \\
\hline
\hline
 \multicolumn{10}{|c|}{electrons} \\
\hline
$c_1$ &1E-16 &1E-15 & 1E-15  &1E-15 &1E-15 &1E-15 &1E-15 &1E-15 &1E-16\\
A & -0.3474  &-0.86156 &-0.8218 &-0.86039 &-0.82019 &-0.8674 &-0.81818
&-0.86596 &-0.325497 \\
B & -0.15596 &-0.03684&-0.04346 &-0.03716 &-0.04412 &-0.0337 &-0.04412 &
-0.03410 & -0.161576\\
\hline
\multicolumn{10}{|c|}{protons} \\
\hline
$c_1$ &1E-16 &1E-15 &1E-15  &1E-15 &1E-15 &1E-15 &1E-15 &1E-15 &1E-15\\
A&-0.3194 &-0.23723*&-0.7393  &-0.2312* &-0.7368 &-0.85357 &-0.7338
&-0.85169 & -0.924767\\
B& 0.1789 &-0.06276*&0.02231  &-0.062768* &0.02238&0.02723  &0.02281
&0.027692 &0.01959\\
\hline
\multicolumn{10}{|c|}{He\ii} \\
\hline
$c_1$ &1E-16 &- &1E-15  &- &1E-15 &1E-16 &1E-15 &1E-16 &1E-16\\
A& -0.5525 &-& -0.84377 & - &-0.84219 &-0.14236 &-0.84037*
&-0.13199&-0.51847 \\
B&0.08565  &-& 0.01192  &- & 0.011451 &0.11010 &0.01177*&0.110894 &
0.092283 \\
\hline
\hline
\multicolumn{10}{|c|}{SHIFT} \\
\hline
\hline
 \multicolumn{10}{|c|}{electrons} \\
\hline
$c_2$ &1E-16 &1E-16 &1E-17  &1E-16 &1E-17 &1E-17  &1E-17 &1E-17 &1E-16\\
 A &-3.4929 &-6.7617 & -81.5107 &-2.5259 & -9.0041
& 49.6736 & -16.5511&40.5533 &23.3420\\
 B &0.01451 &0.00793 & 0.00221  & 0.02206 &0.02144 &0.00526
&0.01131 &0.00586 &0.00261\\
 C &3.3728 &6.6362& 80.9250 & 2.3985 &8.4079 &
-49.60704  &15.9454 &-40.4305 &-23.2800\\
\hline
\multicolumn{10}{|c|}{protons} \\
\hline
$c_2$ &1E-16 &1E-15 &1E-15  &1E-15 &1E-15  &1E-16 &1E-15 & 1E-16 &1E-16\\
 A &-0.0170 &0.25969* &3.0908  &-3.3759* &0.5500 &-2.4538 &
3.7646&18.8373 &  50.1882\\
 B &0.22391& -0.122027* & 0.00215  &  0.00846*&0.00990 &0.02228
&  0.00188&-0.00173 &-0.002557\\
 C &-0.4916 &-0.6076* & -3.2607 &3.0621* & -0.7217&
1.6060 &-3.9381 &-19.6934 &-49.62339\\
\hline
\multicolumn{10}{|c|}{He\ii} \\
\hline
$c_2$ &1E-16 &- &1E-15 &- &1E-15 &1E-16 &1E-15 & 1E-16 &1E-16 \\
 A & -0.0174 &- & 1.7973 &- & 2.4949 &37.4669 &2.0440 &
 36.5920&0.0405\\
 B & 0.20567 &-& 0.01778 &- &0.01314 & 0.00418 &0.01604
&0.00431 &0.22021\\
 C &-0.3771 &- & -1.9474 &- &-2.6462 &-38.2561 &-2.1971
&-37.3904 &0.3832\\
\hline
\end{tabular}
\end{center}
\end{table*}

\subsection{The Stark broadening effect on the shape of Cr\i\, lines} \label{CrB}

In spite of the rather large Stark damping constants the effect is
not observable in stars with solar Cr abundance. In hot stars
where electron and proton densities are high, Cr\i\, lines
considered are generally very weak, while in cooler stars (solar
type) where these lines are strong enough other broadening effects
are more significant. The only chance to look at Stark effect is
the stratified atmosphere of Cr-rich Ap star. The best example is
the well known magnetic star $\beta$~CrB.

Our spectral analysis was based on the spectrum of $\beta$~CrB
obtained
in February 1998 with MuSiCoS spectropolarimeter mounted on the 2
m telescope at Pic du Midi observatory (R=35000). This spectrum
was used by Wade et al. (2001,2003) in stratification analysis.
Starting Cr and Fe distributions were taken from this analysis and
were slightly corrected using a set of spectral lines of Cr\i\,
Cr\ii\, Fe\i\, Fe\ii\ with different excitation energies of the
lower level. The adopted distributions are shown in
Fig.~\ref{strat}. Fe stratification is necessary to account
correctly for blends.

\begin{figure}[h!]
\includegraphics[width=8.5cm]{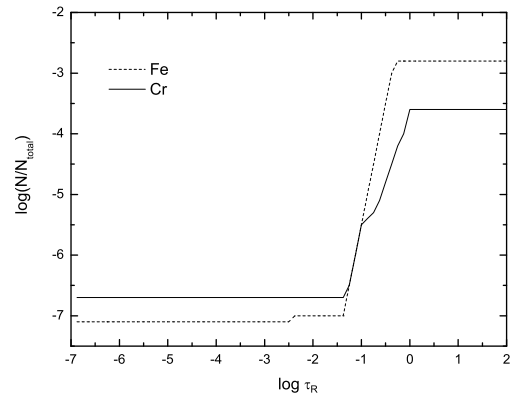}
\caption{Empirically derived Cr and Fe distributions in the atmosphere of
magnetic Ap star $\beta$~CrB.}
 \label{strat}
\end{figure}

\begin{figure}[h!]
\vbox{\vspace{-5mm}
\resizebox{85mm}{!}{{\includegraphics{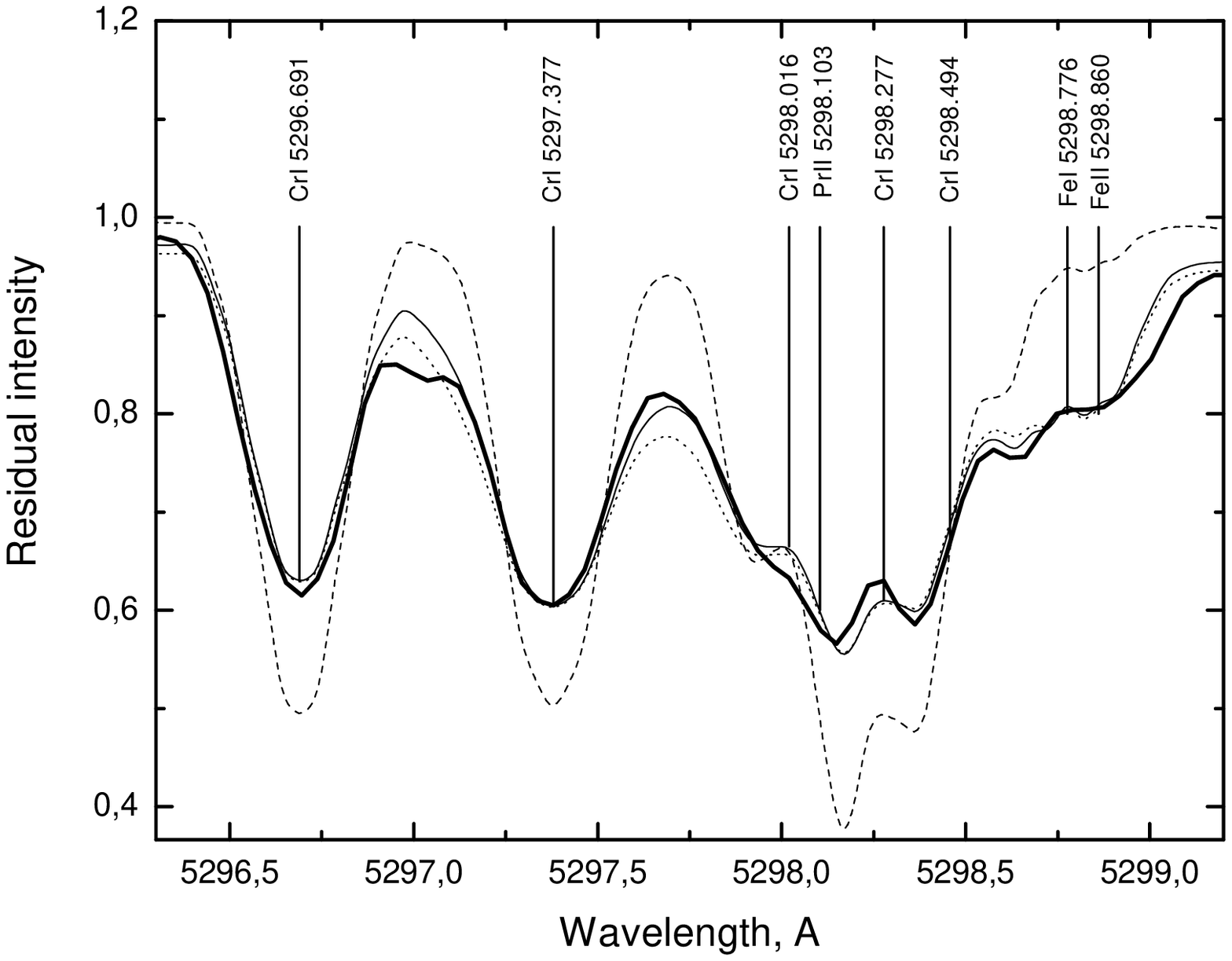}}}}
\vbox{\vspace{-12mm}
\resizebox{85mm}{!}{{\includegraphics{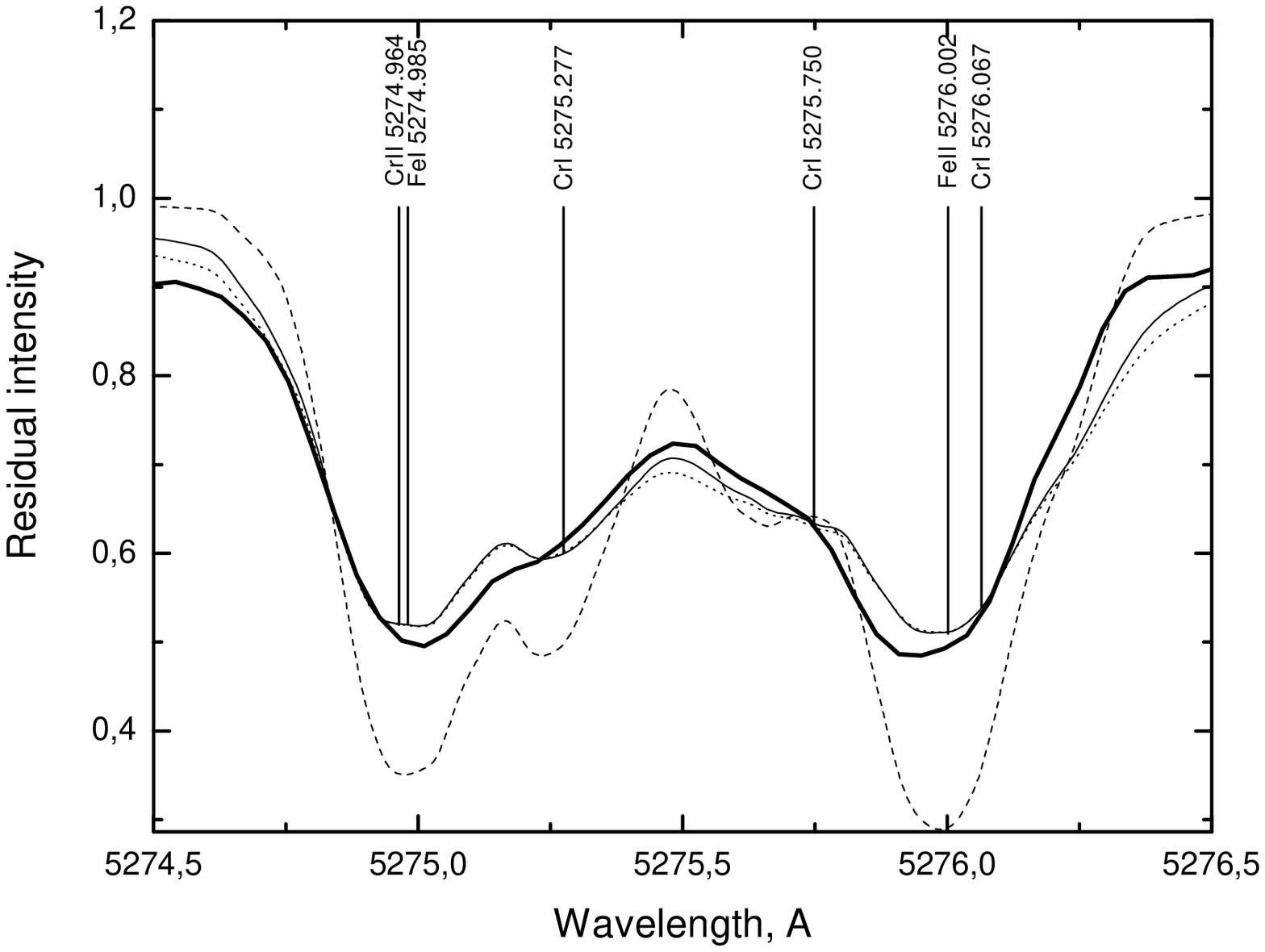}}}}
\vbox{\vspace{-12mm}
\resizebox{85mm}{!}{{\includegraphics{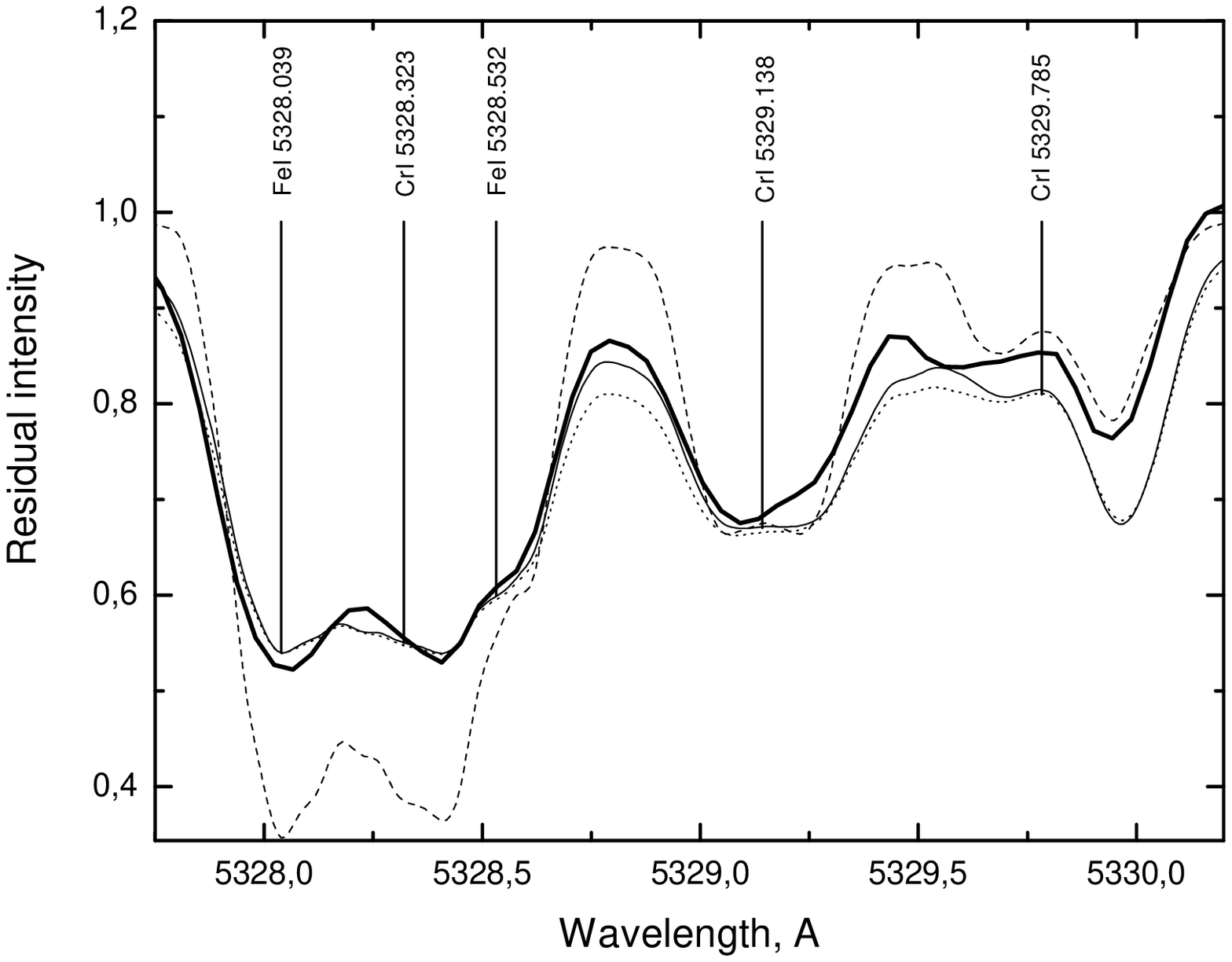}}}} \caption{A
comparison between synthetic spectrum calculations and the
observed spectrum (thick line) of magnetic Ap star $\beta$~CrB in
the regions of Cr\i\, lines 5297 \AA\ (top panel). 5276 \AA\
(middle panel) and 5329 \AA\ (bottom panel). Full thin line -
calculations with stratification shown in Fig~\ref{strat} and
Stark broadening data from Table~\ref{Stark_1} decreased by 70 \%;
dashed line - the same Stark broadening data but homogeneous Cr
and Fe abundances (see text); dotted line - stratified Cr and Fe
abundances, but Stark broadening calculated with the approximation
formula.}
 \label{Cr1}
\end{figure}

Synthetic spectrum calculations in the regions of Cr\i\, lines
considered in the present paper was performed following the method
described in Section~\ref{synthM}. The surface magnetic field
B$_{\rm s}$=5.4 kG  was derived from the magnetically splitted
lines and used in all calculations. In the process of calculations
we found that Cr\i\, lines of our interest have incorrect
wavelengths with the shifts up to 0.05 \AA. We adjusted
wavelengths calculating solar synthetic spectrum and comparing it
with the Solar Flux Atlas (Kurucz et al. 1984). Improved
wavelengths of Cr\i\, lines derived by this way agree within 0.005
\AA\, and better with the precise measurements by J. E. Murray
(1992), kindly provided to us by R.
Kurucz\footnote{http://cfaku5.cfa.harvard.edu/atoms/2400/}. We
placed Murray's wavelengths in Table 1 and 2 and used them in
spectral synthesis. We also estimated the contribution of Van der
Waals broadening and found it to be by the order of magnitude
smaller then Stark broadening through the whole atmosphere above
$log(\tau_{5000})=-0.2$ with the rapid decrease below the
photosphere.

We varied Stark broadening parameters given in Table~\ref{Stark_1}
to achieve the best fit of the calculated Cr\i\, line profiles to
the observations. Fig.~\ref{Cr1} shows a comparison between
observations and calculations in the region of 3 groups of Cr\i\,
lines. For comparison we present synthetic calculations with
homogeneous mean abundances $\log(Cr/N_{tot})=-4.80$ and
$\log(Fe/N_{tot})=-4.15$ which were used in model atmosphere
calculations (Kupka et al. 2004).  To fit Cr\i\, line wings we
need to decrease current Stark widths by 60-70 \%. The same order
of overestimate of Stark damping constants, calculated within the
semiclassical perturbation formalism, was obtained for Si\i\,
lines (see Dimitrijevi\'c et al. 2003). Any change in Cr
distribution, which may be responsible for line profiles, too, is
critical for nearby Cr\i\,$\lambda$~5296.691 \AA, which is
insensitive to Stark broadening mechanism due to small value of
Stark damping constant and is sensitive only to Cr distribution in
stellar atmosphere. We also made a comparison with synthetic
spectrum calculated using stratified Cr distribution and an
approximation formula for Stark broadening (Cowley 1971). This
formula is used in the cases when Stark broadening data,
experimental or theoretical, are not available. Evidently, Stark
broadening  is overestimated when calculations are performed with
this approximation formula.

For detailed investigation of the influence of Stark shifts on the
observed line profile we calculated theoretical spectra with and
without shifts for two separate Cr\i\, lines 5297.377 \AA \- and
5298.494 \AA \- in homogeneous and stratified atmosphere of the
non-rotating star. These two lines have the smallest and largest
shift values, therefore one can make a relative comparison.
Fig.~\ref{shift} illustrates an influence of the Stark shift on
line shape. The total shift in line position due to the Stark
broadening effect is negligible for Cr\i\, 5297.377 \AA\, line
both in homogeneous and stratified atmosphere. For the second line
Stark broadening leads to theoretically measurable shift in the
line position, in particular, in the case of Cr stratification
resulting in a small line asymmetry. Unfortunately, in spectra of
real stars effect is difficult to measure due to other broadening
mechanisms (rotation, magnetic field) and mainly due to severe
blending in the regions with Cr\i\, lines from $4p-4d$
transitions.

\begin{figure}[h!]
\vbox{\vspace{-5mm}
\resizebox{85mm}{!}{\rotatebox{-90}{{\includegraphics{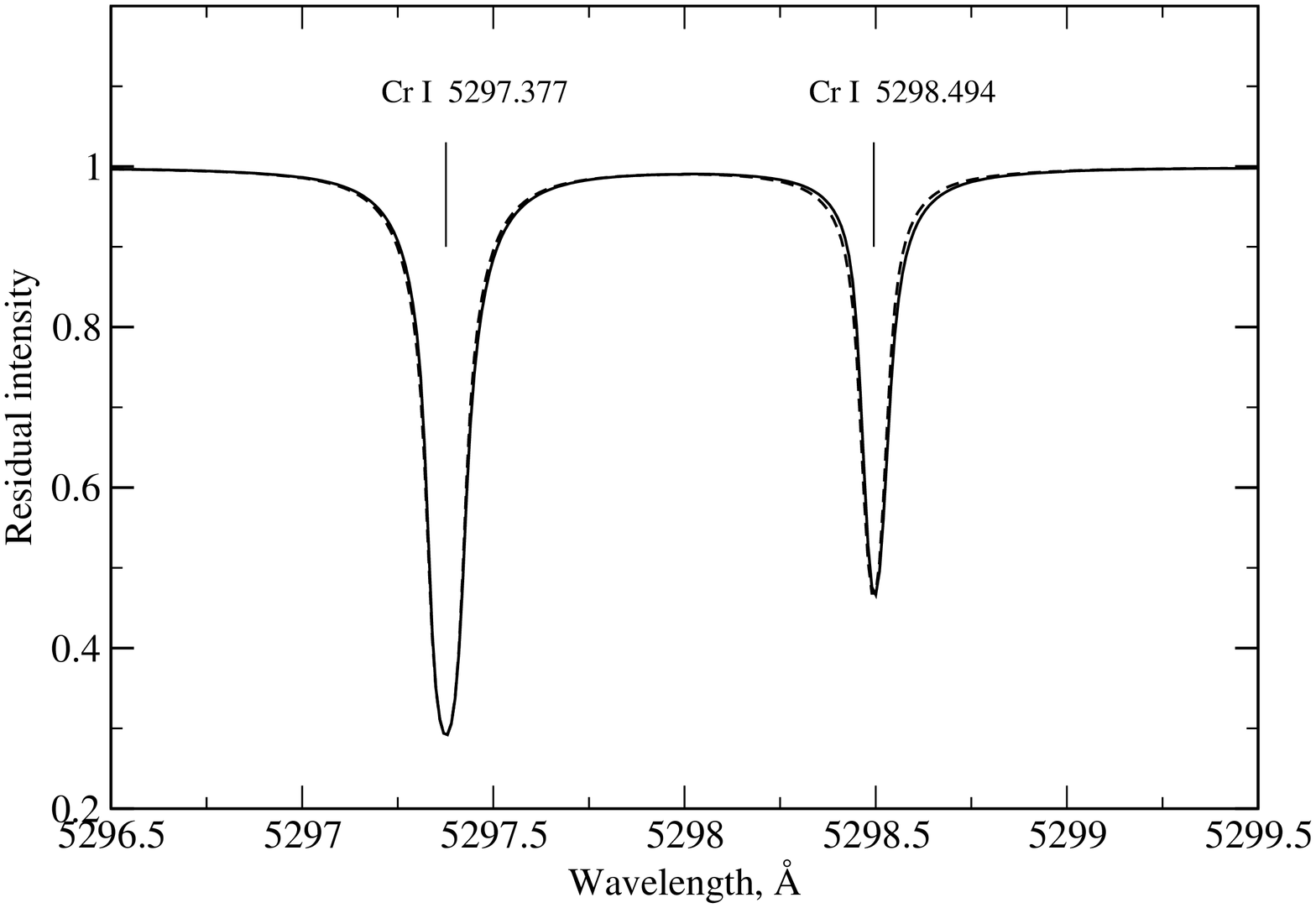}}}}}
\vbox{\vspace{-7mm}
\resizebox{85mm}{!}{\rotatebox{-90}{{\includegraphics{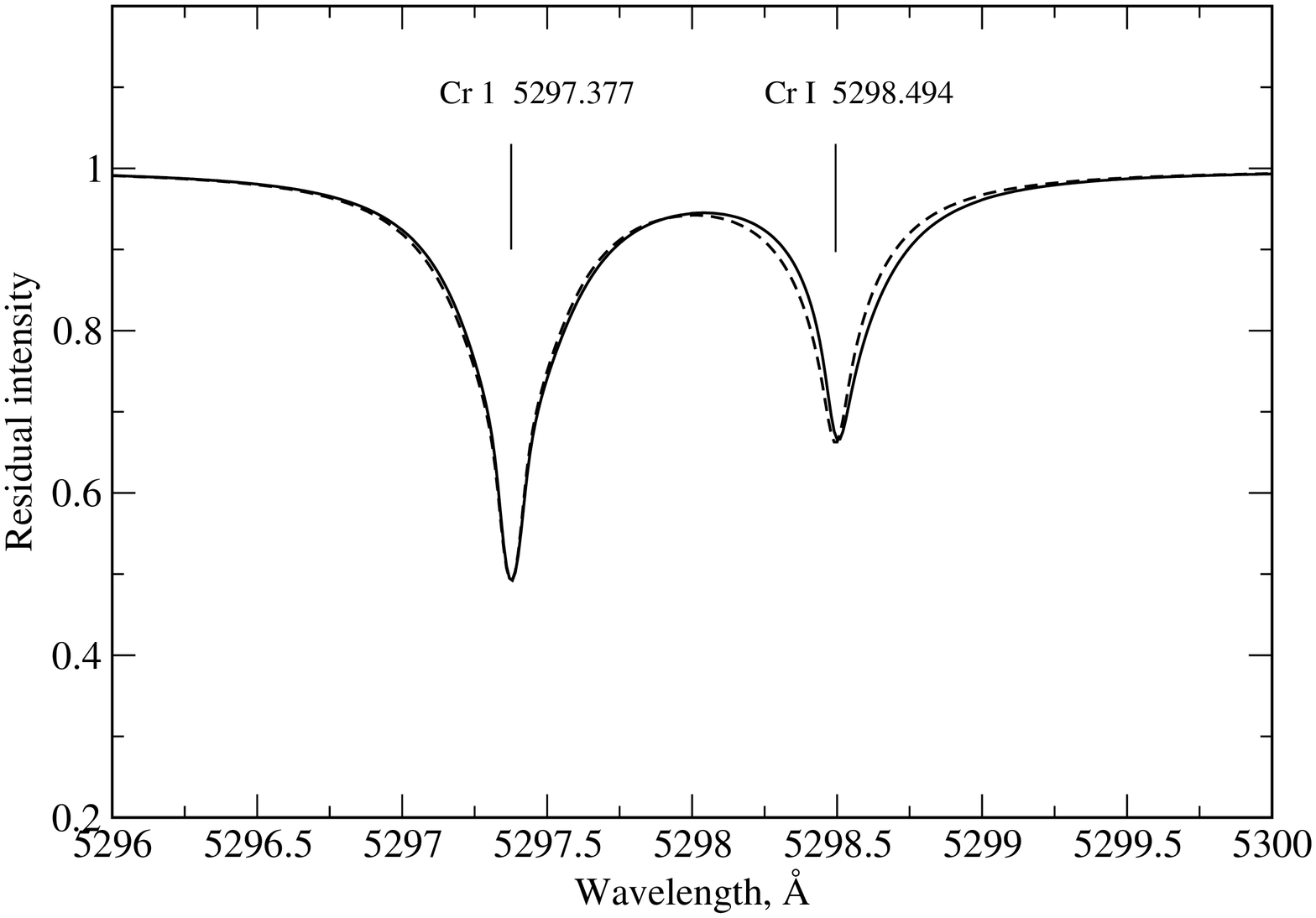}}}}}
\caption{Influence of Stark shift on Cr\i\, line shapes in
homogeneous (upper panel) and in stratified (lower panel)
atmosphere. Calculations with both Stark widths and shifts are
shown by solid lines while those without Stark shifts are shown by
dashed line.}
 \label{shift}
\end{figure}

\section{Conclusions}

Stark broadening parameters for Cr \i\ spectral lines from
$4p^7P^0-4d^7D$ multiplet have been calculated and the influence
of Stark broadening effect in stellar atmosphere for these lines,
has been investigated. From our investigation we can conclude:

(i) Although they belong to a same multiplet, the widths and
shifts of the different lines can be quite different.

(ii) The contribution of the proton and He\ii\ collisions to the
line width and shift is significant, and it is comparable and
sometimes (depending of the electron temperature) even larger than
electron-impact contribution.

(iii) Depending on the electron-, proton-, and He\ii\ density in
stellar atmosphere the Stark shift may contribute to the blue as
well as to the red asymmetry of the same line (see Fig. 2).

(iv) To fit well Cr \i\ line wings we need to decrease the
calculated Stark  widths by 60-70\%, which is the same order of
overestimation as for Si \i\ lines (Dimitrijevi\'c et al. 2003).
The approximation formula of Cowley (1971), used in the cases
where the adequate semiclassical calculation is not possible due
to the lack of reliable atomic data, predicts also overestimated
influence of Stark broadening in comparison with observations.

\begin{acknowledgements}

We are very thankful to G. Wade and R. Kurucz who provided us with
the data necessary for our analysis. This work is a part of the
projects GA-1195``Influence of collisional processes on
astrophysical plasma lineshapes'' and GA-1196 ``Astrophysical
Spectroscopy of Extragalactic Objects'' supported by the Ministry
of Science and Environment protection of Serbia. The research was
supported also by the Fonds zur F\"orderung der wissenschaftlichen
Forschung {\it P14984} . TR thanks Leading Scientific School grant
162.2003.02 and the RFBR (grant 03-02-16342) for partial funding.
L. \v C. P. is supported by Alexander von Humboldt Foundation
through the program for foreign scholars. DS acknowledges
a financial support from INTAS grant 03-55-652.

\end{acknowledgements}

\end{document}